\documentclass[journal=nalefd,manuscript=letter,layout=traditional,email=true]{achemso}

\usepackage{subfigure}
\usepackage{graphicx}
\usepackage{eqnarray}
\usepackage{braket}
\usepackage{graphicx}
\usepackage{natbib}
\usepackage{url}
\usepackage{amsmath}
\usepackage{setspace}
\usepackage{siunitx}
\usepackage{longtable}
\usepackage{mciteplus}
\usepackage{epstopdf}
\usepackage{mhchem}
\usepackage{dsfont}
\usepackage[utf8]{inputenc}

\title{Nanoantenna enhancement for telecom-wavelength superconducting single photon detectors}

\author{Robert M. Heath} %
\email{robert.heath@glasgow.ac.uk} 
\author{Michael G. Tanner} %
\author{Timothy D. Drysdale}
\affiliation{School of Engineering, University of Glasgow, Glasgow, G12 8LT, United Kingdom}

\author{Shigehito Miki} %
\affiliation{Kobe Advanced Research Centre, National Institute of Information and Communications Technology, 588-2, Iwaoka, Iwaoka-cho, Nishi-ku, Kobe, Hyogo, 651-2492, Japan}

\author{Vincenzo Giannini}
\author{Stefan A. Maier}
\affiliation{The Blackett Laboratory, Department of Physics, Imperial College London, London, United Kingdom}

\author{Robert H. Hadfield} %
\affiliation{School of Engineering, University of Glasgow, Glasgow, G12 8LT, United Kingdom}

\begin{document} 
\begin{abstract}
Superconducting nanowire single photon detectors are rapidly emerging as a key infra-red photon-counting technology. Two front-side-coupled silver dipole nanoantennas, simulated to have resonances at \SI{1480}{\nano\metre} and \SI{1525}{\nano\metre}, were fabricated in a two-step process. An enhancement of \SIrange{50}{130}{\percent} in the system detection efficiency was observed when illuminating the antennas. This offers a pathway to increasing absorption into superconducting nanowires, creating larger active areas, and achieving more efficient detection at longer wavelengths.
\end{abstract}


Superconducting nanowire single photon detectors (SNSPDs) are a promising infra-red (IR) single-photon detection technology\cite{hadfield2009single}, with a reported system detection efficiency (SDE) as high as \SI{93}{\percent} at telecommunications wavelengths\cite{marsili2013detecting}. SNSPDs have been trialled in a variety of photon-counting applications including quantum cryptography, ground-to-space communications, and atmospheric remote sensing\cite{natarajan2012superconducting}. Most common approaches to fabricating SNSPDs have a strong electric field polarization dependence\cite{anant2008optical}. This leads to a higher efficiency seen when the incident photons' electric field is polarized in-line with the nanowire. Attempts to mitigate this dependence initially by novel designs using a spiral-geometry meander\cite{dorenbos2008superconducting} or more recently by vertically stacking two efficient WSi meanders in parallel\cite{verma2012three} have offered a route to decreased polarization sensitivity. In contrast to semiconductor single photon detectors there is no sharp cutoff in absorption\cite{baek2011superconducting,marsili2012efficient}; thus SNSPDs have considerable potential for use at mid-IR wavelengths\cite{korneeva2011new}. 

Optical antennas at visible and IR wavelengths couple incident photons into the local plasmon resonance of the antenna material\cite{biagioni2012nanoantennas}, allowing strong localization of the electric field below the optical limit\cite{aizpurua2005optical}. Some recent work has been carried out to enhance SNSPD performance by exploiting plasmonics: one approach has been to deposit a full layer of gold over the nanowires and couple in through the back side of the substrate\cite{hu2011superconducting} exploiting non-resonant collection and resonant cavity behaviour, while another approach used nanocavities and reflectors rather than nanoantennas\cite{csete2013improvement}. In this work, we seek to demonstrate conclusively that absorption of incident IR light in a superconducting nanowire can be enhanced via plasmon resonance in a silver dipole nanoantenna. This could allow for SNSPDs with a sparse meander and larger active area that maintain the favorable efficiency and timing properties of a smaller device. 

\begin{figure}[]
	\setlength\fboxsep{0pt}
	\setlength\fboxrule{0.25pt}
	\begin{center}
	    \includegraphics[width=0.9\textwidth]{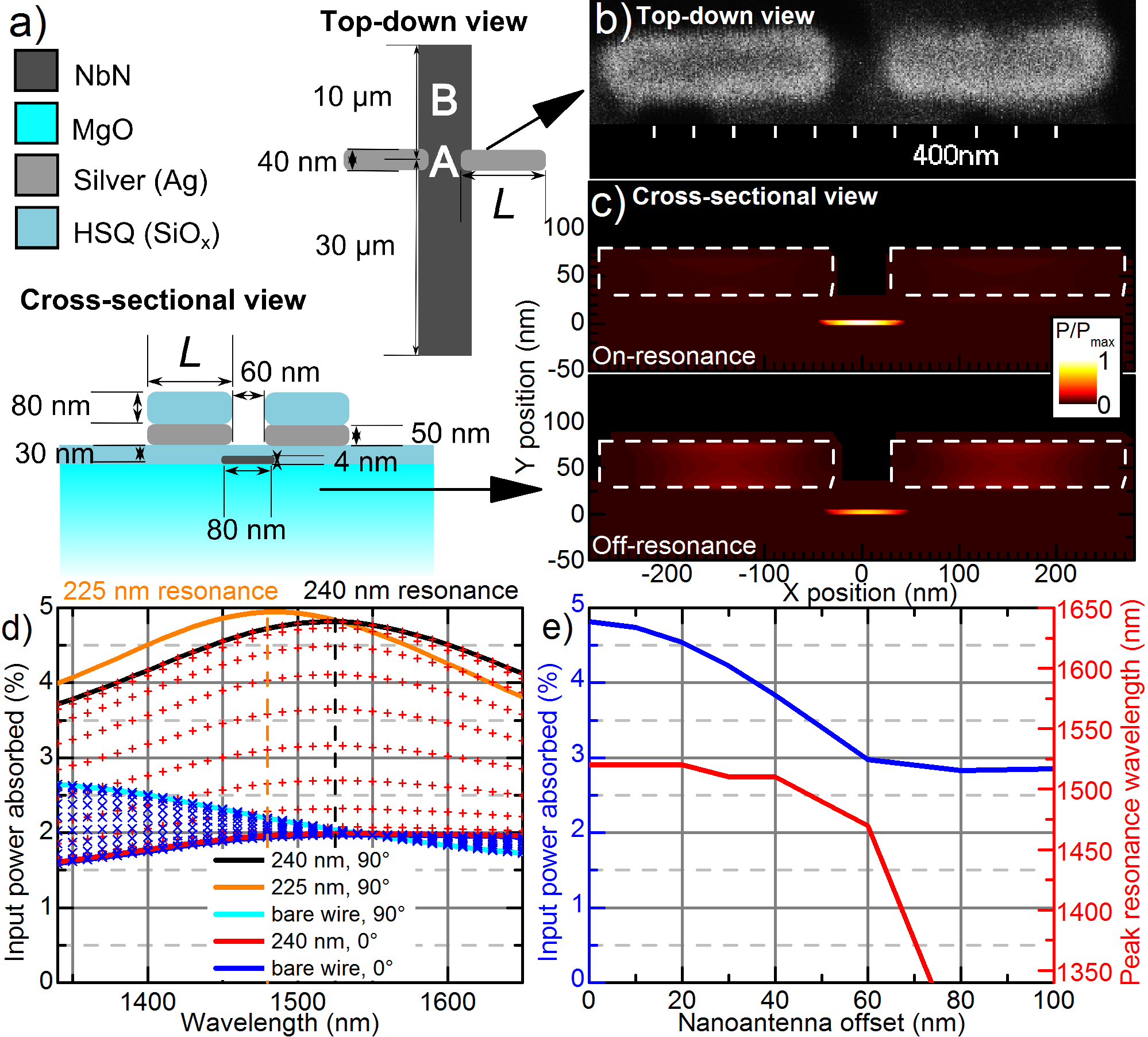}
	\end{center}
	\caption{Figure~\ref{fig:sem-laser}a outlines the simulation setup used to model the power absorbed into the nanowire (diagram not to scale). Figure~\ref{fig:sem-laser}b is a top-down SEM image showing the $L_{act}=$~\SI{240}{\nano\metre} nanoantenna. The nanowire is not visible due to being buried under \SI{30}{\nano\metre} \ce{SiO_{x}}. After etching, residual HSQ resist forms a thin layer of \ce{SiO_{x}} on the surface. Figure~\ref{fig:sem-laser}c shows a normalized slice of the simulation of the absorbed power for the $L_{act}=$~\SI{240}{\nano\metre} antennna, on-resonance (\SI{1520}{\nano\metre}) and off-resonance (\SI{1000}{\nano\metre}), the antennas marked with white dashed lines.  The output of simulation of the total absorbed power into the nanowire is shown in figure~\ref{fig:sem-laser}d across a range of linear electric field polarizations. Red pluses represent the  $L_{act}=$~\SI{240}{\nano\metre} antenna simulation data, and are bounded by a black line and a red line to denote the perpendicular to the nanowire (\SI{90}{\degree}) and parallel to the nanowire (\SI{0}{\degree}) polarization responses to incident light. Blue crosses represent the bare nanowire photoresponse, bounded by a cyan line and a blue line to denote \SI{90}{\degree} and \SI{0}{\degree} polarizations. Additionally, the \SI{90}{\degree} polarization response for the $L_{act}=$~\SI{225}{\nano\metre} antenna is shown in orange for later comparison, and the resonances of the antennas are marked with dashed lines. Determining the effect of offsetting the Ag nanoantenna from its location centrally over the NbN nanowire is shown in figure~\ref{fig:sem-laser}e. This model treats the detection efficiency as simply the ratio of the incident power to the absorbed power in the nanowire.}
	\label{fig:sem-laser}
\end{figure}

A \SI{4}{\nano\metre} thick film of niobium nitride (\ce{NbN}) was deposited by reactive DC magnetron sputtering at room temperature\cite{miki2007nbn,miki2008large} on a magnesium oxide (\ce{MgO}) substrate and patterned into a \SI{40}{\micro\metre}-long, \SI{80}{\nano\metre}-wide nanowire. \SI{30}{\nano\metre} of silicon oxide (\ce{SiO_{x}}) and \SI{50}{\nano\metre} of silver (Ag) were then deposited. The nanoantennas were patterned by e-beam lithography into hydrogen silsesquioxane (HSQ) and etched with argon (\ce{Ar}) ion milling. Finally, an \ce{SiO_{x}} capping layer was deposited. The nanoantenna geometry comprised of two \SI{40}{\nano\metre} wide nanoantenna halves fabricated \SI{60}{\nano\metre} apart, with a design length per half of $L_{des}=$~\SI{220}{\nano\metre} or $L_{des}=$~\SI{210}{\nano\metre} (see figure~\ref{fig:sem-laser}a). These have a dominant scattering cross-section\cite{giannini2011plasmonic}. These dimensions were chosen as they offered the best compromise of enhancement and ease of fabrication. No adhesion layer was used for the silver as this would have reduced the absorption enhancement. The scanning electron microscope (SEM) image shown in figure~\ref{fig:sem-laser}b suggests that the nanoantenna with a design length of $L_{des}=$~\SI{220}{\nano\metre} became slightly longer after patterning, with a measured length of $L_{act}=$~\SI{240}{\nano\metre}, with the $L_{des}=$~\SI{210}{\nano\metre} antenna becoming $L_{act}=$~\SI{225}{\nano\metre}. Simulations were performed using $L_{act}$ rather than the design length, to better match with the experimental result.

The incident power absorbed into the nanowire was modelled using the finite-difference time-domain (FDTD) method in Lumerical FDTD Solutions. The simulation parameters are shown in figure~\ref{fig:sem-laser}a. These simulations modelled the power absorbed at each point. This allows a cut-through `slice' of the absorbed power both at the wavelength where the absorbed power into the nanowire is a maximum (referred to as on-resonance) and at wavelengths other than this (referred to as off-resonance), shown in figure~\ref{fig:sem-laser}c. On-resonance it is clear that the majority of the power is absorbed into the nanowire, peaking at $1.00$ in the middle of the nanowire; off-resonance, the antenna absorbs a significant amount, and the peak power in the nanowire reaches only $0.72$. Additionally, the total power absorbed into the entire length of the nanowire was obtained from the simulation, and was the most readily comparable metric with the SNSPD's detection efficiency.

\begin{figure}[]
	\setlength\fboxsep{0pt}
	\setlength\fboxrule{0.25pt}
	\begin{center}
		\includegraphics[width=0.9\textwidth]{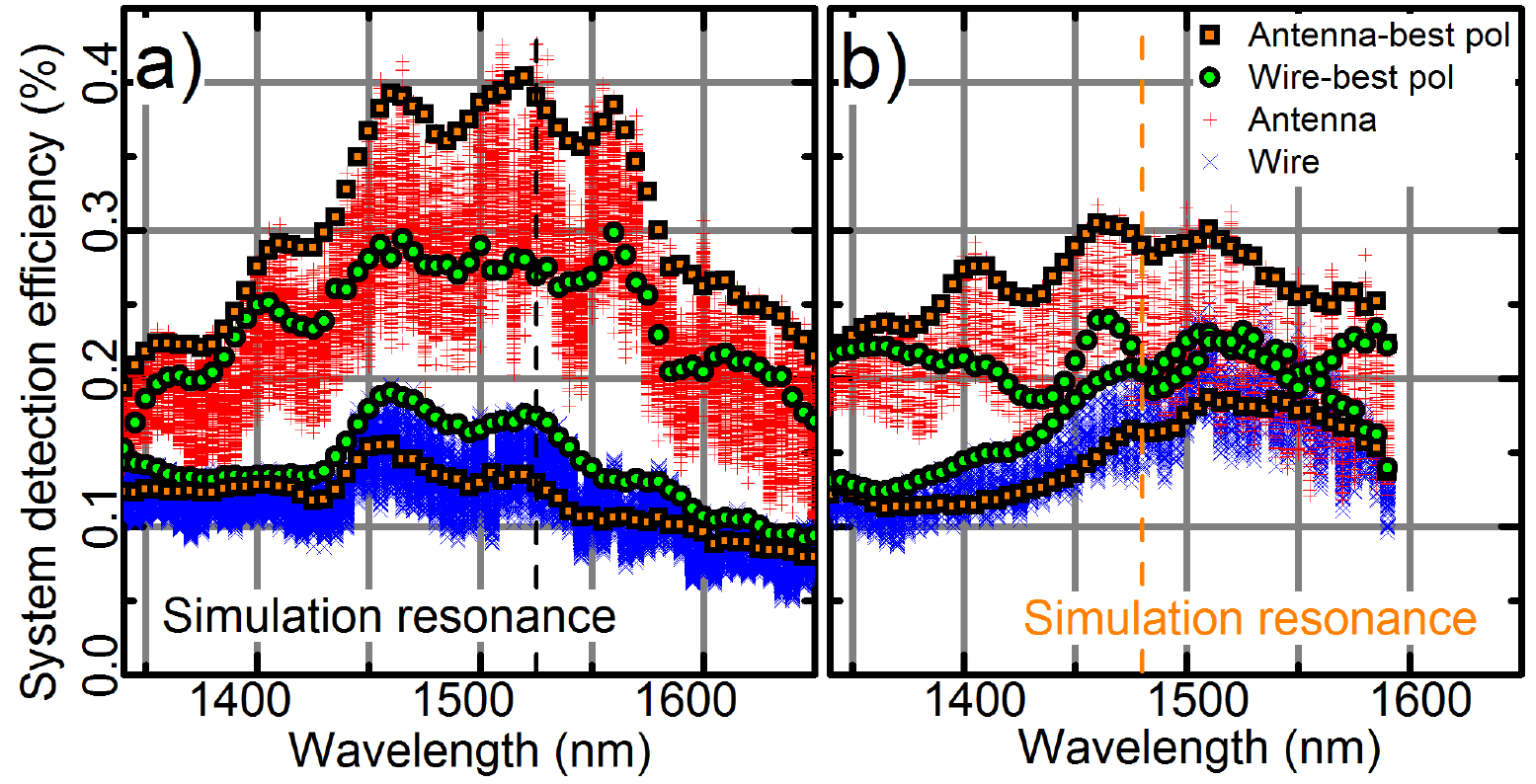}
		
	    
	\end{center}
	\caption{By sweeping the wavelength from \SI{1340}{\nano\metre} to \SI{1650}{\nano\metre} in \SI{5}{\nano\metre} steps and normalizing out the photon flux using a separate power measurement it is possible to observe the range of efficiencies at each wavelength across the full polarization space. By using an electronic polarization controller these polarizations are made repeatable, which allows post-measurement selection of efficiencies at a given polarization across multiple separate measurements. In both figures, red pluses represent data obtained when illuminating an antenna, while blue crosses represent data obtained illuminating a bare wire. In figure~\ref{fig:results-figure}a the polarizations at which the $\lambda_{res}=$~\SI{1525}{\nano\metre} nanoantenna and wire respond best are selected. Figure~\ref{fig:results-figure}b shows the response for the $\lambda_{res}=$~\SI{1480}{\nano\metre} nanoantenna. The peak value data for these figures was averaged over the four adjacent points to smooth out the interference. There is clear agreement between the peak efficiency value and the simulation's suggested resonance (shown as a dashed line, comparable with figure~\ref{fig:sem-laser}d). Figure~\ref{fig:results-figure}a sampled 256 polarizations per wavelength while figure~\ref{fig:results-figure}b sampled 64: separate testing showed no improvement using 256 points over 64, and it was quicker to measure fewer polarizations.} 
	\label{fig:results-figure}
\end{figure}

The FDTD simulation shown in figure~\ref{fig:sem-laser}d shows the expected antenna and bare wire behaviour for the $L_{act}=$~\SI{240}{\nano\metre} antennae while rotating the linear electric field polarizations from \SI{0}{\degree} (in line along the wire) to \SI{90}{\degree} (perpendicular to the wire) in steps of \SI{10}{\degree}. This data suggested a resonance wavelength $\lambda_{res}$ for the $L_{act}=$~\SI{240}{\nano\metre} nanoantenna of $\lambda_{res}=$~\SI{1525}{\nano\metre}, while the $L_{act}=$~\SI{225}{\nano\metre} nanoantenna's simulated resonance wavelength was $\lambda_{res}=$~\SI{1480}{\nano\metre}. The crossing of the polarization sensitivity in the `no antenna' case is attributable to an unavoidable numerical artifact: a \SI{13}{\percent} reflection from the \ce{SiO_{x}} spacer and the simulation boundary reaches a resonance condition at \SI{1540}{\nano\metre}, enhancing absorption in all polarisations and obscuring the true polarization sensitivity of the wire. We explored the parameter space to ensure this artifact is understood and small enough not to significantly influence the results with an antenna, and thus can be discounted, by comparing different sizes of simulation. It is estimated that a small wire of these dimensions could have a degree of polarization of less than \SI{100}{\percent}, but is too small to become fully unpolarized\cite{agdur1963scattering}. This is consistent with experimental observations (figure~\ref{fig:results-figure}).




The simulations were employed to assess expected behaviour if the nanowire was misaligned with the nanoantenna, shown in figure~\ref{fig:sem-laser}e for the $L_{act}=$~\SI{240}{\nano\metre} nanoantenna. Due to the HSQ cap and the low profile of the nanowire it was not possible to resolve the nanowire in the SEM. For a small misalignment the peak resonance wavelength (red) is pushed to lower wavelengths and the absorbed power decreases (blue), until a misalignment of \SI{80}{\nano\metre}, when the peak resonance moves below \SI{1340}{\nano\metre}, the lowest wavelength of our tunable lasers. Subsequently, we can determine that if our experimental setup is able to characterize the resonance peak, the upper bound on the misalignment of the nanowire and nanoantenna is less than \SI{80}{\nano\metre}.

It was important to control the electric field polarization during this experiment as both nanowires and nanoantennas are polarization-sensitive. Measuring the electric field incident on the nanowire in-situ is non-trivial as the device is at \SI{3.5}{\kelvin}, so a relative measurement scale is used: a computer-controlled polarizer is deterministically swept across a number of positions of the Poincaré sphere and the `best' and `worst' count rates are used as positions of reference. From literature one might expect the `wire best' polarization to be a linear polarization in-line with the wire (\SI{0}{\degree} in simulated data), and `wire worst' to be a polarization perpendicular to the wire\cite{anant2008optical} (\SI{90}{\degree} in simulated data) but these were observed for a meander geometry SNSPD, which is fundamentally a diffraction grating, and so may not necessarily apply here. Simulation suggests that the power absorbed into the nanowire peaks when polarization is in-line with the antenna (\SI{90}{\degree}), and that the antenna behaves worst when the polarization is in-line with the wire (\SI{0}{\degree}, which would not excite an antenna resonance). We are reluctant to assert that these are necessarily `antenna best' and `antenna worst' without a more rigorous study connecting detection efficiency and absorbed power into the nanowire. Furthermore, the simulation was limited to linear polarizations, and so ignored elliptical and circular polarizations. Subsequently, the `best' and `worst' labels are used only in reference to what is experimentally observed.

As described below, in order to cover a large spectral range a system of three tunable lasers is employed. The electric field polarizations cannot be assumed to be maintained as the lasers change and each laser sweeps wavelength. For this reason, at every wavelength, a scan through the electric field polarization space was taken, and the `wire best' and `antenna best' polarizations were extracted retrospectively. As the polarization sweeps were made using a computer-controlled polarizer it was possible to compare the exact response of two spatially-distinct regions of the detector, allowing comparisons for each polarization from both the antenna and the bare wire. The locations used are shown in figure~\ref{fig:sem-laser}a as A and B.

The detector tested had a \SI{9}{\micro\ampere} critical current when cooled to \SI{3.5}{\kelvin} in a vibration-damped cryostat based on a closed-cycle Cryomech PT403-RM pulse-tube cryocooler. Light was delivered by fiber, to a miniature confocal microscope configuration with a spot of \SI{1.3}{\micro\metre}\cite{oconnor2011spatial}, and was used to illuminate the nanoantenna detectors from the front side. The wavelength of the input light was varied from \SI{1340}{\nano\metre} to \SI{1650}{\nano\metre}, a spectral range of \SI{310}{\nano\metre}, in \SI{5}{\nano\metre} steps.

Light was inserted from a combination of an Agilent HP 8168F (range \SI{1440}{\nano\metre} to \SI{1590}{\nano\metre}) and two modules in a Yenista OSICS rack: an ECL-1400 (range \SI{1340}{\nano\metre} to \SI{1440}{\nano\metre}) and T100-1620 (range \SI{1560}{\nano\metre} to \SI{1680}{\nano\metre}) through an electronically-controlled optical switch. This was then passed through two Agilent 8156A variable optical attenuators connected in series to reduce the input photon flux, and was then passed through an Agilent 11896A electronic polarization controller. At this point, to calibrate the incident light, the power was measured across the range with \SI{0}{\deci\bel} applied attenuation at the input to the cryostat. The power meter, a Thorlabs PM 300, was disconnected and the fiber connected to the cryostat's optical input. The detector response was tested, and afterwards the laser power profile was re-tested to confirm the stability of the laser power. The power meter and attenuators have calibrations traceable to the National Institute of Standard and Technology.

A power variation of \SI{3}{\deci\bel} was observed across the wavelength range of the lasers. The varying power of the laser was normalized out afterward: the applied attenuation and laser power were summed and the photon flux calculated. From this it was possible to determine the SDE at each wavelength and electric field polarization.The optical attenuators were calibrated between \SI{1340}{\nano\metre} and \SI{1650}{\nano\metre}, so all measurements were within this range.

Looking at the detector response when the wavelength of the incident light is varied we are able to observe the nanoantenna resonance in the photoresponse of the detector. There are some oscillations in the SDE with wavelength, but as others have noted\cite{baek2011superconducting} these are an interference effect. In this case, the spacing of the oscillations suggests a characteristic length of about \SI{75}{\micro\metre} which does not clearly correspond to any designed characteristic length, but as noted the data in figures~\ref{fig:results-figure}a and~\ref{fig:results-figure}b have been given a 5-point moving average, and the data was only taken every \SI{5}{\nano\metre}, which means this could be an alias of a larger characteristic length.

\begin{figure}[]
	\setlength\fboxsep{0pt}
	\setlength\fboxrule{0.25pt}
	\begin{center}
		\includegraphics[width=0.9\textwidth]{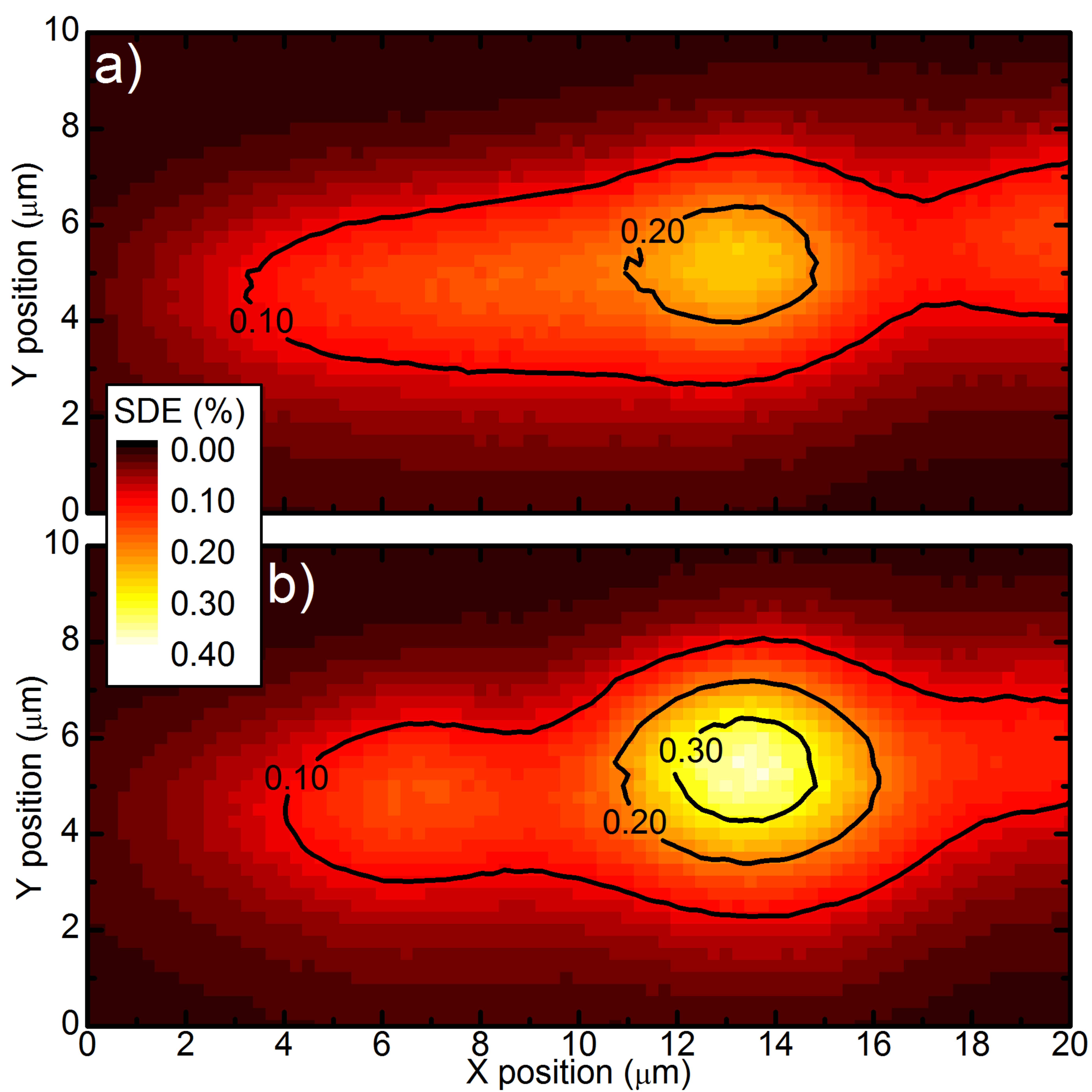}
	\end{center}
	\caption{By raster-scanning the miniature confocal microscope over the active area of the SNSPD, maps of the photodetection profile are created. Figure~\ref{fig:photoresponse}a shows a map of the nanowire when the incident light is polarized with the electric field such that the antenna responds least, the so-called `antenna worst' polarization. Figure~\ref{fig:photoresponse}b is the same plot with the polarization rotated to give the highest response from the antenna, the `antenna best' polarization. There is significant improvement on `antenna worst' response at `antenna best', and it is clear that in both cases the antenna has a positive effect on the detection rate.}
	\label{fig:photoresponse}
\end{figure}

Figure~\ref{fig:results-figure}a shows the \SI{1525}{\nano\metre}-resonant nanoantenna's response and the response from the bare wire. As the settings of the polarization controller are recorded it is possible to link one dataset's `best' response with the response of the other dataset at that same polarization: for this reason, there are two `antenna best' and `wire best' lines plotted. At the `antenna best' polarization, there is strong enhancement over the response of the bare wire at the same polarization: furthermore, the `antenna best' polarization and the `wire best' are different, which is expected as both the nanoantenna and nanowire are polarization-sensitive, and these are perpendicular. The enhancement at the resonant wavelength is \SI{130}{\percent}, the SDE increasing from \SI{0.17}{\percent} to \SI{0.40}{\percent}. Looking at figure~\ref{fig:results-figure}b, which is the \SI{1480}{\nano\metre}-resonant nanoantenna response and response from bare wire, the resonance wavelength seen, as simulation suggested, is lower, but the enhancement is less strong, going from an SDE of \SI{0.20}{\percent} to \SI{0.30}{\percent}, an increase of \SI{50}{\percent}. Whether this is due to poor fabrication or poor alignment with the nanowire is unclear, but the effect is still a net improvement the photoresponse at the design wavelength. 

Using the low-temperature piezoelectric miniature confocal microscope employed in previous work\cite{heath2014nano,casaburi2014parallel}, we map the photoresponse of the \SI{1525}{\nano\metre}-resonant detector with sub-micrometer precision. By varying the polarization we are able to spatially visualise the polarization sensitivity of the detector. Figure~\ref{fig:photoresponse} shows photoresponse from the nanowire and antenna with different polarizations of incident light. The nanowire runs horizontally through the image, with the antenna at about \SI{10}{\micro\metre} along the wire's photoresponse. In figure~\ref{fig:photoresponse}a we see the `antenna worst' polarization: the antenna has only a small effect, boosting the wire response of \SI{0.15}{\percent} to \SI{0.25}{\percent}. Rotating the polarization to the `antenna best' polarization, we obtain figure~\ref{fig:photoresponse}b: the nanoantenna responds much more strongly, enhancing the detection efficiency from \SI{0.15}{\percent} to \SI{0.37}{\percent}. This qualitatively agrees with the FDTD simulation of the absorbed power into the nanowire: divergence in absolute values likely stems from the simulation reporting the ratio of absorbed power into the nanowire, which is not necessarily directly comparable with the detection efficiency\cite{natarajan2012superconducting}.

Two nanoantennas with different resonant wavelengths were successfully fabricated atop superconducting nanowires and were comprehensively characterized at low temperature. From comparing the position of the photoresponse resonance with simulation we can conclude that the nanoantennas had less than \SI{80}{\nano\metre} of misalignment with the nanowire. The nanoantennas increased absorption into the nanowire, so creating a localised region of increased SDE with a separate polarization sensitivity, and offered a \SIrange{50}{130}{\percent} improvement in the local SDE. Unlike other nanoantenna-coupled SNSPD designs\cite{hu2011superconducting}, this does not suppress the SDE at any tested electric field polarization, at the expense of less improvement at the preferential polarization. This approach offers a route to improving SNSPD efficiency at longer wavelengths and could be employed to create larger-area detectors with a reduced fill factor. Fabrication for longer wavelengths would also be less challenging due to the structures being physically larger. It may also offers a route to reducing the polarization sensitivity of meander structures. Future work may include increasing the number of antennas and decreasing the gap between the antennas to increase enhancement, as well as tailoring performance for mid-infrared wavelengths.

During final preparation of this manuscript it has come to our attention that preliminary data on near-IR antennas coupled to superconducting nanowires was reported by R. Heeres\cite{reinierheeresthesis}. The authors thank the UK Engineering and Physical Sciences Research Council (EPSRC) for support. The authors also thank Saburo Imamura of the Japanese National Institute of Information and Communications Technology (NICT) for technical support. RMH thanks NICT for an Internship Research Fellowship. SAM and VG acknowledge support from the Leverhulme Trust and the EPSRC Active Plasmonics Programme. RHH acknowledges a Royal Society of London University Research Fellowship. The authors declare no competing financial interest.

\bibliography{library}

\providecommand{\latin}[1]{#1}
\providecommand*\mcitethebibliography{\thebibliography}
\csname @ifundefined\endcsname{endmcitethebibliography}
  {\let\endmcitethebibliography\endthebibliography}{}
\begin{mcitethebibliography}{22}
\providecommand*\natexlab[1]{#1}
\providecommand*\mciteSetBstSublistMode[1]{}
\providecommand*\mciteSetBstMaxWidthForm[2]{}
\providecommand*\mciteBstWouldAddEndPuncttrue
  {\def\EndOfBibitem{\unskip.}}
\providecommand*\mciteBstWouldAddEndPunctfalse
  {\let\EndOfBibitem\relax}
\providecommand*\mciteSetBstMidEndSepPunct[3]{}
\providecommand*\mciteSetBstSublistLabelBeginEnd[3]{}
\providecommand*\EndOfBibitem{}
\mciteSetBstSublistMode{f}
\mciteSetBstMaxWidthForm{subitem}{(\alph{mcitesubitemcount})}
\mciteSetBstSublistLabelBeginEnd
  {\mcitemaxwidthsubitemform\space}
  {\relax}
  {\relax}

\bibitem[Hadfield(2009)]{hadfield2009single}
Hadfield,~R.~H. \emph{Nat. Photonics} \textbf{2009}, \emph{3}, 696--705\relax
\mciteBstWouldAddEndPuncttrue
\mciteSetBstMidEndSepPunct{\mcitedefaultmidpunct}
{\mcitedefaultendpunct}{\mcitedefaultseppunct}\relax
\EndOfBibitem
\bibitem[Marsili \latin{et~al.}(2013)Marsili, Verma, Stern, Harrington, Lita,
  Gerrits, Vayshenker, Baek, Shaw, Mirin, and Nam]{marsili2013detecting}
Marsili,~F.; Verma,~V.~B.; Stern,~J.~A.; Harrington,~S.; Lita,~A.~E.;
  Gerrits,~T.; Vayshenker,~I.; Baek,~B.; Shaw,~M.~D.; Mirin,~R.~P.; Nam,~S.~W.
  \emph{Nat. Photonics} \textbf{2013}, \emph{7}, 210--214\relax
\mciteBstWouldAddEndPuncttrue
\mciteSetBstMidEndSepPunct{\mcitedefaultmidpunct}
{\mcitedefaultendpunct}{\mcitedefaultseppunct}\relax
\EndOfBibitem
\bibitem[Natarajan \latin{et~al.}(2012)Natarajan, Tanner, and
  Hadfield]{natarajan2012superconducting}
Natarajan,~C.~M.; Tanner,~M.~G.; Hadfield,~R.~H. \emph{Supercond. Sci.
  Technol.} \textbf{2012}, \emph{25}, 063001\relax
\mciteBstWouldAddEndPuncttrue
\mciteSetBstMidEndSepPunct{\mcitedefaultmidpunct}
{\mcitedefaultendpunct}{\mcitedefaultseppunct}\relax
\EndOfBibitem
\bibitem[Anant \latin{et~al.}(2008)Anant, Kerman, Dauler, Yang, Rosfjord, and
  Berggren]{anant2008optical}
Anant,~V.; Kerman,~A.~J.; Dauler,~E.~A.; Yang,~J. K.~W.; Rosfjord,~K.~M.;
  Berggren,~K.~K. \emph{Opt. Express} \textbf{2008}, \emph{16},
  10750--10761\relax
\mciteBstWouldAddEndPuncttrue
\mciteSetBstMidEndSepPunct{\mcitedefaultmidpunct}
{\mcitedefaultendpunct}{\mcitedefaultseppunct}\relax
\EndOfBibitem
\bibitem[Dorenbos \latin{et~al.}(2008)Dorenbos, Reiger, Akopian, Perinetti,
  Zwiller, Zijlstra, and Klapwijk]{dorenbos2008superconducting}
Dorenbos,~S.~N.; Reiger,~E.~M.; Akopian,~N.; Perinetti,~U.; Zwiller,~V.;
  Zijlstra,~T.; Klapwijk,~T.~M. \emph{Appl. Phys. Lett.} \textbf{2008},
  \emph{93}, 161102\relax
\mciteBstWouldAddEndPuncttrue
\mciteSetBstMidEndSepPunct{\mcitedefaultmidpunct}
{\mcitedefaultendpunct}{\mcitedefaultseppunct}\relax
\EndOfBibitem
\bibitem[Verma \latin{et~al.}(2012)Verma, Marsili, Harrington, Lita, Mirin, and
  Nam]{verma2012three}
Verma,~V.~B.; Marsili,~F.; Harrington,~S.; Lita,~A.~E.; Mirin,~R.~P.;
  Nam,~S.~W. \emph{Appl. Phys. Lett.} \textbf{2012}, \emph{101}, 251114\relax
\mciteBstWouldAddEndPuncttrue
\mciteSetBstMidEndSepPunct{\mcitedefaultmidpunct}
{\mcitedefaultendpunct}{\mcitedefaultseppunct}\relax
\EndOfBibitem
\bibitem[Baek \latin{et~al.}(2011)Baek, Lita, Verma, and
  Nam]{baek2011superconducting}
Baek,~B.; Lita,~A.~E.; Verma,~V.; Nam,~S.~W. \emph{Appl. Phys. Lett.}
  \textbf{2011}, \emph{98}, 251105\relax
\mciteBstWouldAddEndPuncttrue
\mciteSetBstMidEndSepPunct{\mcitedefaultmidpunct}
{\mcitedefaultendpunct}{\mcitedefaultseppunct}\relax
\EndOfBibitem
\bibitem[Marsili \latin{et~al.}(2012)Marsili, Bellei, Najafi, Dane, Dauler,
  Molnar, and Berggren]{marsili2012efficient}
Marsili,~F.; Bellei,~F.; Najafi,~F.; Dane,~A.~E.; Dauler,~E.~A.; Molnar,~R.~J.;
  Berggren,~K.~K. \emph{Nano Lett.} \textbf{2012}, \emph{12}, 4799--4804\relax
\mciteBstWouldAddEndPuncttrue
\mciteSetBstMidEndSepPunct{\mcitedefaultmidpunct}
{\mcitedefaultendpunct}{\mcitedefaultseppunct}\relax
\EndOfBibitem
\bibitem[Korneeva \latin{et~al.}(2011)Korneeva, Florya, Semenov, Korneev, and
  Gol'tsman]{korneeva2011new}
Korneeva,~Y.; Florya,~I.; Semenov,~A.; Korneev,~A.; Gol'tsman,~G.~N. \emph{IEEE
  Trans. Appl. Supercond.} \textbf{2011}, \emph{21}, 323--326\relax
\mciteBstWouldAddEndPuncttrue
\mciteSetBstMidEndSepPunct{\mcitedefaultmidpunct}
{\mcitedefaultendpunct}{\mcitedefaultseppunct}\relax
\EndOfBibitem
\bibitem[Biagioni \latin{et~al.}(2012)Biagioni, Huang, and
  Hecht]{biagioni2012nanoantennas}
Biagioni,~P.; Huang,~J.-S.; Hecht,~B. \emph{Rep. Prog. Phys.} \textbf{2012},
  \emph{75}, 024402\relax
\mciteBstWouldAddEndPuncttrue
\mciteSetBstMidEndSepPunct{\mcitedefaultmidpunct}
{\mcitedefaultendpunct}{\mcitedefaultseppunct}\relax
\EndOfBibitem
\bibitem[Aizpurua \latin{et~al.}(2005)Aizpurua, Bryant, Richter, Garc{\'\i}a~de
  Abajo, Kelley, and Mallouk]{aizpurua2005optical}
Aizpurua,~J.; Bryant,~G.~W.; Richter,~L.~J.; Garc{\'\i}a~de Abajo,~F.~J.;
  Kelley,~B.~K.; Mallouk,~T. \emph{Phys. Rev. B} \textbf{2005}, \emph{71},
  235420\relax
\mciteBstWouldAddEndPuncttrue
\mciteSetBstMidEndSepPunct{\mcitedefaultmidpunct}
{\mcitedefaultendpunct}{\mcitedefaultseppunct}\relax
\EndOfBibitem
\bibitem[Hu \latin{et~al.}(2011)Hu, Dauler, Molnar, and
  Berggren]{hu2011superconducting}
Hu,~X.; Dauler,~E.~A.; Molnar,~R.~J.; Berggren,~K.~K. \emph{Opt. Express}
  \textbf{2011}, \emph{19}, 17--31\relax
\mciteBstWouldAddEndPuncttrue
\mciteSetBstMidEndSepPunct{\mcitedefaultmidpunct}
{\mcitedefaultendpunct}{\mcitedefaultseppunct}\relax
\EndOfBibitem
\bibitem[Csete \latin{et~al.}(2013)Csete, Sipos, Szalai, Najafi, Szab{\'o}, and
  Berggren]{csete2013improvement}
Csete,~M.; Sipos,~A.; Szalai,~A.; Najafi,~F.; Szab{\'o},~G.; Berggren,~K.~K.
  \emph{Sci. Rep.} \textbf{2013}, \emph{3}, 2406\relax
\mciteBstWouldAddEndPuncttrue
\mciteSetBstMidEndSepPunct{\mcitedefaultmidpunct}
{\mcitedefaultendpunct}{\mcitedefaultseppunct}\relax
\EndOfBibitem
\bibitem[Miki \latin{et~al.}(2007)Miki, Fujiwara, Sasaki, and
  Wang]{miki2007nbn}
Miki,~S.; Fujiwara,~M.; Sasaki,~M.; Wang,~Z. \emph{IEEE Trans. Appl.
  Supercond.} \textbf{2007}, \emph{17}, 285--288\relax
\mciteBstWouldAddEndPuncttrue
\mciteSetBstMidEndSepPunct{\mcitedefaultmidpunct}
{\mcitedefaultendpunct}{\mcitedefaultseppunct}\relax
\EndOfBibitem
\bibitem[Miki \latin{et~al.}(2008)Miki, Fujiwara, Sasaki, Baek, Miller,
  Hadfield, Nam, and Wang]{miki2008large}
Miki,~S.; Fujiwara,~M.; Sasaki,~M.; Baek,~B.; Miller,~A.~J.; Hadfield,~R.~H.;
  Nam,~S.~W.; Wang,~Z. \emph{Appl. Phys. Lett.} \textbf{2008}, \emph{92},
  061116\relax
\mciteBstWouldAddEndPuncttrue
\mciteSetBstMidEndSepPunct{\mcitedefaultmidpunct}
{\mcitedefaultendpunct}{\mcitedefaultseppunct}\relax
\EndOfBibitem
\bibitem[Giannini \latin{et~al.}(2011)Giannini, Fernandez-Dominguez, Heck, and
  Maier]{giannini2011plasmonic}
Giannini,~V.; Fernandez-Dominguez,~A.~I.; Heck,~S.~C.; Maier,~S.~A. \emph{Chem.
  Rev.} \textbf{2011}, \emph{111}, 3888--3912\relax
\mciteBstWouldAddEndPuncttrue
\mciteSetBstMidEndSepPunct{\mcitedefaultmidpunct}
{\mcitedefaultendpunct}{\mcitedefaultseppunct}\relax
\EndOfBibitem
\bibitem[Agdur \latin{et~al.}(1963)Agdur, B{\"o}ling, Sellberg, and
  {\"O}hman]{agdur1963scattering}
Agdur,~B.; B{\"o}ling,~G.; Sellberg,~F.; {\"O}hman,~Y. \emph{Phys. Rev.}
  \textbf{1963}, \emph{130}, 996\relax
\mciteBstWouldAddEndPuncttrue
\mciteSetBstMidEndSepPunct{\mcitedefaultmidpunct}
{\mcitedefaultendpunct}{\mcitedefaultseppunct}\relax
\EndOfBibitem
\bibitem[O'Connor \latin{et~al.}(2011)O'Connor, Tanner, Natarajan, Buller,
  Warburton, Miki, Wang, Nam, and Hadfield]{oconnor2011spatial}
O'Connor,~J.~A.; Tanner,~M.~G.; Natarajan,~C.~M.; Buller,~G.~S.;
  Warburton,~R.~J.; Miki,~S.; Wang,~Z.; Nam,~S.~W.; Hadfield,~R.~H. \emph{Appl.
  Phys. Lett.} \textbf{2011}, \emph{98}, 201116\relax
\mciteBstWouldAddEndPuncttrue
\mciteSetBstMidEndSepPunct{\mcitedefaultmidpunct}
{\mcitedefaultendpunct}{\mcitedefaultseppunct}\relax
\EndOfBibitem
\bibitem[Heath \latin{et~al.}(2014)Heath, Tanner, Casaburi, Webster, San
  Emeterio~Alvarez, Jiang, Barber, Warburton, and Hadfield]{heath2014nano}
Heath,~R.~M.; Tanner,~M.~G.; Casaburi,~A.; Webster,~M.~G.; San
  Emeterio~Alvarez,~L.; Jiang,~W.; Barber,~Z.~H.; Warburton,~R.~J.;
  Hadfield,~R.~H. \emph{Appl. Phys. Lett.} \textbf{2014}, \emph{104},
  063503\relax
\mciteBstWouldAddEndPuncttrue
\mciteSetBstMidEndSepPunct{\mcitedefaultmidpunct}
{\mcitedefaultendpunct}{\mcitedefaultseppunct}\relax
\EndOfBibitem
\bibitem[Casaburi \latin{et~al.}(2014)Casaburi, Heath, Tanner, Cristiano,
  Ejrnaes, Nappi, and Hadfield]{casaburi2014parallel}
Casaburi,~A.; Heath,~R.~M.; Tanner,~M.~G.; Cristiano,~R.; Ejrnaes,~M.;
  Nappi,~C.; Hadfield,~R.~H. \emph{Supercond. Sci. Technol.} \textbf{2014},
  \emph{27}, 044029\relax
\mciteBstWouldAddEndPuncttrue
\mciteSetBstMidEndSepPunct{\mcitedefaultmidpunct}
{\mcitedefaultendpunct}{\mcitedefaultseppunct}\relax
\EndOfBibitem
\bibitem[Heeres(2012)]{reinierheeresthesis}
Heeres,~R. {Quantum Plasmonics}. Ph.D.\ thesis, {Technische Universiteit
  Delft}, {The Netherlands}, 2012\relax
\mciteBstWouldAddEndPuncttrue
\mciteSetBstMidEndSepPunct{\mcitedefaultmidpunct}
{\mcitedefaultendpunct}{\mcitedefaultseppunct}\relax
\EndOfBibitem
\end{mcitethebibliography}

\end{document}